\documentclass{svjour3}
\smartqed
\usepackage{amsmath}
\usepackage{graphicx}
\usepackage{float}

\begin{document}

\title{Thermodynamics of Charged Lovelock - AdS Black Holes}
\author{C.B Prasobh\and Jishnu Suresh\and V.C Kuriakose}
\institute{C.B Prasobh\at Department of Physics, Cochin University of Science and Technology, Cochin 682202, India
\email{mail.prasobhcb@gmail.com}
\and Jishnu Suresh\at Department of Physics, Cochin University of Science and Technology, Cochin 682202, India
\email{jishnusuresh@cusat.ac.in}
\and V.C Kuriakose\at Department of Physics, Cochin University of Science and Technology, Cochin 682202, India
\email{vck@cusat.ac.in}
}
\date{Received: date / Accepted: date}
\maketitle

\begin{abstract}
We investigate the thermodynamic behavior of maximally symmetric charged, asymptotically AdS black hole solutions of Lovelock gravity. We explore the thermodynamic stability of such solutions by the ordinary method of calculating the specific heat of the black holes and investigating its  divergences which signal second order phase transitions between black hole states. We then utilize the methods of thermodynamic geometry of black hole spacetimes in order to explain the origin of these points of divergence. We calculate the curvature scalar corresponding to a Legendre-invariant thermodynamic metric of these spacetimes and find that the divergences in the black hole specific heat correspond to singularities in the thermodynamic phase space. We also calculate the area spectrum for large black holes in the model by applying the Bohr-Sommerfeld quantization to the adiabatic invariant calculated for the spacetime.


\PACS{PACS code1 \and PACS code2 \and more}

\end{abstract}

\section{Introduction}\label{sec:intro}

The subject of black hole thermodynamics had its origin in the observation\cite{hawking1,hawking2,hawking3,hawking4,bekenstein,bardeen,penrose} of a mathematical connection between various quantities that are relevant to black hole dynamics - horizon-area, mass, surface gravity etc. and thermodynamic variables - entropy, temperature etc. that describe the thermodynamic behaviour of systems. Consequences of this mathematical connection drive the current intense activity in this field, more than four decades after its initial discovery. We now suspect that this connection actually goes much deeper than a simple one-to-one correspondence between various parameters. It is known that many aspects of quantum field theories of various systems have their dual in gravitational systems. This connection enables us to analyze the behaviour of such systems by studying their dual gravitational theories, which is often a much easier task. Recently discovered gauge-gravity dualities like the AdS/CFT correspondence\cite{maldacena}, according to which asymptotically AdS gravitational theories in $d$ dimensions are dual to quantum field theories in a $d-1$ dimensional sub-manifold, have fueled intense interest in asymptotically AdS spacetimes, which started after it was pointed out that thermal radiation / large AdS black hole phase transitions can take place\cite{page}. 

\paragraph{}


The occurrence of phase transitions between various black hole states is a very important aspect of thermodynamic studies of gravitational systems, since it would enable us to study the behaviour of their dual systems near their critical points. These phase transitions can be studied in various ways - studying the heat capacity of black hole spacetimes is one approach\cite{cvetic_gubser,caldarelli}, in which the positivity of the specific heat would point to a stable phase of the black hole while a negative value signals an unstable phase. Transitions between thermal AdS space and black hole configurations, discovered by Hawking\cite{page}, is considered as the pioneering study on the subject. According to it, pure thermal radiation in AdS space becomes unstable above a certain temperature and collapses to form black holes. This is the well-known Hawking-Page phase transition which describes the phase transition between the Schwarzschild AdS black hole and the thermal AdS space. This is dual\cite{witten} to the confinement/deconfinement phase transition of gauge fields according to the AdS/CFT correspondence\cite{maldacena}. Since then, phase transitions of black holes have been investigated from different perspectives. Some recent works may be found in \cite{Sahay,Banerjee1,Banerjee2,Cao,Banerjee3,Banerjee4,Banerjee5,Banerjee6,Weishaowen1,Majhi,Kim1,Tsai,Capela,Kubiznak,Niu,Lala,Lala2,Weishaowen2,Eune,Poshteh,Wenbiao1}. 

\paragraph{}

Another approach to analyze black hole thermodynamic stability is to apply the methods of differential geometry by considering the thermodynamic phase space of a black hole system as a Riemannian manifold and studying its curvature, which would then represent thermodynamic interaction\cite{rao,Alvarez,gibbs,hermann,mrugala1,mrugala2,weinhold,ruppeiner,quevedo1,quevedo2,quevedo3,quevedo4}. This curvature is determined by assigning a metric to the thermodynamic phase space. The components of the metric are defined in terms of second derivatives of suitable thermodynamic potentials with respect to a set of extensive variables $N^a$ of the thermodynamic system. Usual choices for the thermodynamic potentials are the mass $M$, internal energy $U$, entropy $S$, etc. of the black hole spacetime. Depending on the choice of the metric, different versions of the geometric approach exist. The thermodynamic geometry method was first introduced by Weinhold\cite{weinhold} and Ruppeiner \cite{ruppeiner}. Weinhold proposed a metric structure in the energy representation as $g_{ij}^{W}=\partial_{i}\partial_{j}\ M(U,N^a)$, while Ruppeiner defined the metric structure as $g_{ij}^{R}=-\partial_{i}\partial_{j}\ S(U,N^a)$. Components of these metrics are those of the Hessian matrix of the internal energy $M$ and the entropy $S$ respectively, with respect to the extensive thermodynamic variables $N^a$. Weinhold's metric was found to be conformally connected to Ruppeiner's through the relation $ds^2_R=\dfrac{ds^2_W}{T}$\cite{Janyszek}, $T$ being the horizon temperature. Ruppeiner's metric has extensively been used in the geometric analysis of various black hole spacetimes\cite{Ruppeiner2}. Recently, Quevedo et al.\cite{quevedo2} presented a new formalism called geometrothermodynamics, which allows us to derive Legendre invariant metrics for the phase space. Geometrothermodynamics presents a unified geometry where the metric structure describes various types of black hole thermodynamics \cite{quevedo1,quevedo2,quevedo3,quevedo4,quevedo5,quevedo6,jishnu1,jishnu2,tharanath1,tharanath2}.

\paragraph{}

Theoretical interest in the black hole horizon area stems from arguments\cite{strominger,ashtekar} that the origin of horizon entropy is related to the quantum structure of spacetime. Statistical mechanics tells us that entropy is a measure of the number of occupied microstates of a system that have equal probability of being occupied. The direct counting of these microstates in the case of black hole spacetimes is still an unresolved problem. On one hand, entropy must obey the second law of thermodynamics, according to which it can do nothing but increase. On the other hand, we also know from the no-hair-theorem that the state of the black hole systems must be specified by a mere handful of parameters, namely the mass $M$, the charge $Q$ and the angular momentum $L$ of the black hole. In other words, a large portion of information regarding the fields that collapse to form black holes get lost to the observable universe, so that the nature of the microstates becomes obscure. This leads to a violation of unitarity since, according to quantum mechanics, pure states can only evolve into pure states, whereas the state inside the black hole becomes mixed after its formation. There have been suggestions\cite{hsu} that gravitational collapse could lead to the formation of topologically disconnected regions where the information could be stored. Thus, black hole horizon area is considered to be intimately related to the very process of black hole formation and could offer vital glimpses into the quantum nature of spacetime itself, and thus be of incredible help in the formulation of a quantum theory of gravity. Following the initial proposal of Bekenstein\cite{bekenstein1,bekenstein2,bekenstein3} of the discrete nature of the black hole spectrum, various approaches have been developed for the computation of the same\cite{hod1,hod2,dreyer,maggiore,kunstatter,setare1,setare2,setare3,setare4,medved,vagenas}.

\paragraph{}

In the present work, we test the thermodynamic stability of black holes in charged, asymptotically AdS, spherically symmetric spacetimes in Lovelock model. Ordinary thermodynamic analysis reveals the existence of two points in charged spacetimes, where the specific heat as a function of the horizon radius diverges, compared to just one in the uncharged case. Then we compute the scalar curvature of the thermodynamic phase space for the spacetime using a Legendre-invariant metric proposed by Quevedo\cite{quevedo1} and find that there exist divergences in the scalar curvature near the points of divergence of the specific heat, thus explaining the thermodynamic phase transitions. We then calculate the area spectrum of black hole horizons in the model by directly calculating the adiabatic invariant for the spacetime and applying the Bohr-Sommerfeld quantization condition to it. A brief outline of the paper is as follows: in Sect. \ref{sect:ord_thermo}, we explain the maximally symmetric Lovelock model and the resulting metric for the charged AdS black hole spacetime\cite{bhscan}. We also calculate the relevant thermodynamic quantities like the horizon temperature, entropy and the specific heat in the same section. Details of the geometrothermodynamic method of analyzing the phase transitions are given in Sect. \ref{sect:gtd}. Calculation of the adiabatic invariant for the spacetime and the deduction of the area spectrum of large black hole are performed in Sect. \ref{sect:area}. The results are summarized in Sect. \ref{sect:conclusion}.

\section{Thermodynamic stability of charged AdS black holes in Lovelock model}\label{sect:ord_thermo}

The Lovelock model of gravity\cite{lovelock,wheeler} is developed based on a Lagrangian in the form of a polynomial in the Riemann curvature. The degree of the polynomial determines the order of the resulting theory. It is known\cite{takahashithirdorder,takahashieven,takahashiflatunstable} that the stability of the solutions to these theories against metric perturbations is not always guaranteed. The black hole spacetimes, whose thermodynamic stability is studied in the present work, are solutions to a subset of the general Lovelock theories, restricted by the additional constraint that all the solutions must possess a unique AdS vacuum state with a fixed cosmological constant\cite{bhscan}. In such theories, the order $k$ of the corresponding Lagrangian labels the different theories and it is seen that the type of the theory depends on the values of $k$ and  dimension $d$ of the spacetime. For $d>3$, the metric representing the spherically symmetric, charged, asymptotically AdS solutions to such theories, is given by\cite{bhscan},

\begin{equation}\label{metric}
ds^{2}=f(r)dt^{2}+\frac{dr^{2}}{f(r)}+r^{2}d\Omega_{d-2}^{2}\ ,
\end{equation}

where $f(r)$ is given by

\begin{equation}\label{f}
f(r)=1+\frac{r^{2}}{R^{2}}-g(r),
\end{equation}

where $g(r)=\Biggl[\dfrac{2G_kM}{r^{d-2k-1}}-\left(\dfrac{\epsilon G_k}{d-3}\right)\dfrac{Q^2}{r^{2(d-k-2)}}\Biggr]^{\dfrac{1}{k}}$. Here, $r$ is a Schwarzschild-like coordinate and $R$ is the unique AdS radius, related to the cosmological constant $\Lambda$ by the relation $\Lambda=\dfrac{(d-1)(d-2)}{2R^{2}}$. The value of $R$ is taken to be equal to 1 for all the numerical calculations in this paper. $G_k$ refers to the gravitational constant for the theory of order $k$. The constant $\epsilon$ is proportional to the permeability of the vacuum in $d$ dimensions. The constants $M$ and  $Q$ refer to the mass and the electric charge of the black hole respectively. It is also known\cite{bhscan} that there exist lower limits for the mass of the black hole $M$ and the size $r_e$ of the charged object, as long as we wish to avoid time-like singularities.

\paragraph{}

\begin{figure}[H]
\begin{center}
\includegraphics[scale=0.75]{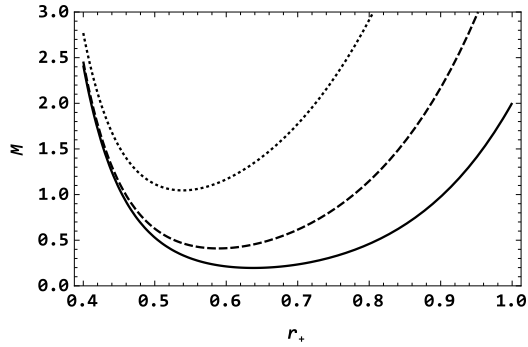}
\caption{\label{fig:M_vs_rplus_Q_0p235_d10_k_2_bottom_to_4_top}Mass of the charged black hole as a function of $r_+$. The curves are drawn for $d=10$ and $Q=0.235$, with  $k=2$ (solid), $k=3$ (dashed) and $k=4$ (dotted).}
\end{center}
\end{figure}

The event horizon $r_+$ of the black hole is taken as the largest positive root of the equation $f(r)=0$. For arbitrary values of the parameters $d$ and $k$, it is obviously not possible to express $r_+$ as a function of the parameters $M,~Q,~R,$ etc. However, it is possible to express the mass $M$ of the black hole as a function of $r_+$, which is plotted in Fig. \ref{fig:M_vs_rplus_Q_0p235_d10_k_2_bottom_to_4_top}. The function $M$ is expressed in terms of $r_+$ as,

\begin{equation}\label{mass}
M(r_+)=\dfrac{r_+^{d-2k-1}}{2G_k}\Biggl[\left(1+\dfrac{r_+^2}{R^2}\right)^k+\left(\dfrac{\epsilon G_k}{d-3}\right)\dfrac{Q^2}{r_+^{2(d-k-2)}}\Biggr].
\end{equation}

The horizon temperature $T$ is obtained by requiring the Euclidean time to be periodic with period $\tau =4\pi \left( \left. \dfrac{df}{dr}\right| _{r=r_{+}}\right) ^{-1}$ and equating it to $\dfrac{1}{\kappa_B T}$, $\kappa_B$ being the Boltzmann constant. We can easily see that, for the charged black holes, the horizon temperature is given by

\begin{equation}\label{temperature}
T(r_+)=\frac{1}{4\pi \kappa _{B}}\left. \frac{df}{dr}\right| _{r=r_+}=\dfrac{2r_+}{R^2}-\dfrac{1}{k}\Biggl[g(r_+)\Biggr]^{\dfrac{1}{k}-1}\times g'(r_+).
\end{equation}

\begin{figure}[H]
\begin{center}
\includegraphics[scale=0.75]{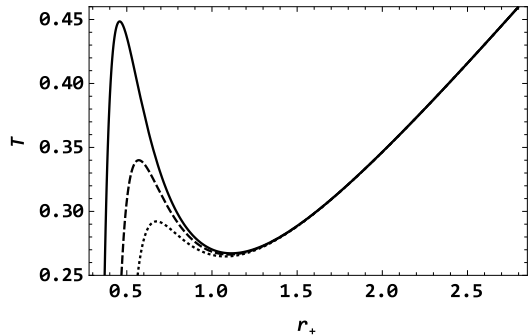}
\caption{\label{fig:T_vs_rplus_d7k2_Q_0p2_top_to_0p4_bottom}Temperature of the charged black hole as a function of $r_+$, for $d=7$, $k=2$. The curves are drawn for $Q=0.2$ (solid), $Q=0.3$ (dashed) and $Q=0.4$ (dotted).}
\end{center}
\end{figure}

Eq. (\ref{temperature}) represents a non-monotonic function having a couple of turning points when expressed as a function of $r_+$. Once again, it is not possible to obtain closed-form expressions for these points in terms of the black hole parameters, as long as the dimension $d$ and order $k$ are not fixed. However, one can analyze the behaviour of $T(r_+)$ as a function of $r_+$ graphically. Fig. \ref{fig:T_vs_rplus_d7k2_Q_0p2_top_to_0p4_bottom} represents a plot between $T(r_+)$ and $r_+$. From the plot, it is obvious that one of the turning points represents a maximum while the other is a minimum. The existence of a minimum of $T(r_+)$ is known already in the case of uncharged black holes\cite{bhscan}, while the existence of the maximum appears to be unique to the charged case. 

\paragraph{}

The function $S(r_+)$, representing the entropy of the black hole event horizon as a function of the horizon radius $r_+$ is obtained in the general case of a Lovelock theory of order $k$ at spacetime dimension $d$, is found out by evaluating the Euclidean action for the back hole spacetime and equating it to $\beta$ times the free energy of the system, where $\beta$ is defined by the expression $\left. \dfrac{df}{dr}\right|_{r_+}=4\pi \beta ^{-1}$. After some calculation, it is seen that the entropy $S(r_+)$ is given by,

\begin{equation}\label{entropy}
S(r_+)=\dfrac{r_+^{d-2 k}}{d-2 k}\ _2F_1\left(\dfrac{1}{2} (d-2 k),1-k;\dfrac{1}{2} (d-2 k+2);-\dfrac{r_+^2}{R^2}\right)
\end{equation} 

\begin{figure}[H]
\begin{center}
\includegraphics[scale=0.75]{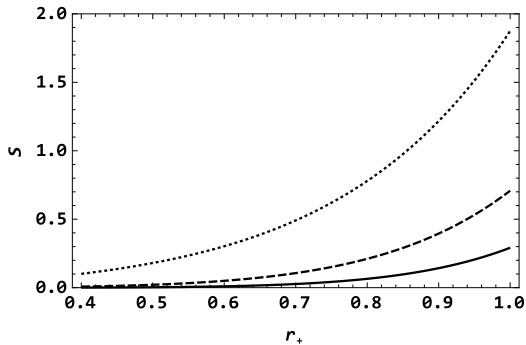}
\caption{\label{fig:S_vs_rplus_Q_0p235_d10_k_2_bottom_to_4_top}Entropy of the charged black hole as a function of $r_+$. The curves are drawn for $d=10$ and $Q=0.235$, with  $k=2$ (solid), $k=3$ (dashed) and $k=4$ (dotted).}
\end{center}
\end{figure}

where $_2F_1$ represents the hypergeometric function. Graphically, it is seen from Fig. \ref{fig:S_vs_rplus_Q_0p235_d10_k_2_bottom_to_4_top} that $S(r_+)$ is a monotonically increasing function of $r_+$. For $k=1$, it can readily be seen that $S(r_+) \propto r_+^{d-2}$, which is nothing but the usual area law, namely $S \propto A$, $\ A$ being the horizon-area. Also, $S(r_+)$ becomes proportional to the area for $k \neq 1$ theories when $r_+ \gg R$, i.e. for very large black holes.

\paragraph{}

In order to investigate the thermodynamic stability of the black hole spacetime, we compute the specific heat $C_p$ for the spacetime, defined as $C_p=\dfrac{\partial M}{\partial T}$. Since both $M$ and $T$ can conveniently be expressed as functions of $r_+$, we compute $C_p$ also as a function of $r_+$ using the expression $C_p(r_+)=\dfrac{\partial_{r_+} M}{\partial_{r_+} T}$. The explicit form of the function turns out to be too long to include here, so that we resort to graphical analysis.

\paragraph{}

\begin{figure}[H]
\begin{center}
\includegraphics[scale=1.0]{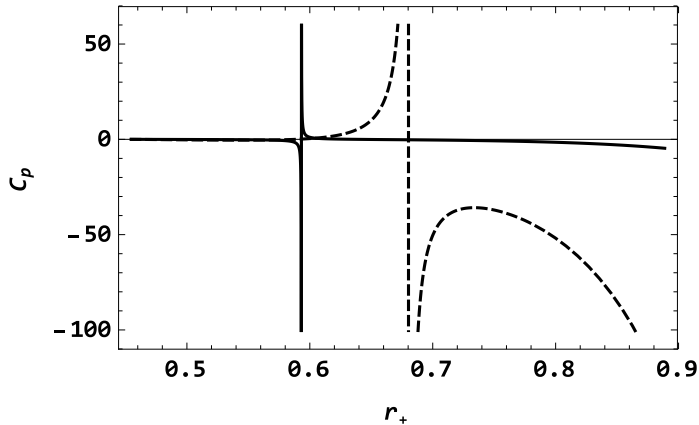}
\caption{\label{fig:C_vs_rplus_rc1_Q_0p235_d10_k_2_solid_k_3_dashed}$C_p~-~r_+$ variation in the vicinity of $r_{c1}$. The curves are drawn with $d=10$, $k=2$ (solid) and $d=10$, $k=3$ (dashed). In both cases, $Q=0.235$}
\end{center}
\end{figure}

\begin{figure}[H]
\begin{center}
\includegraphics[scale=1.0]{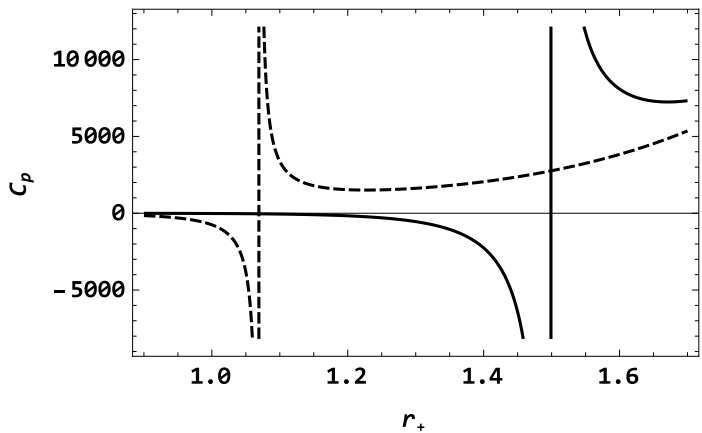}
\caption{\label{fig:C_vs_rplus_rc2_Q_0p235_d10_k_2_solid_k_3_dashed}$C_p~-~r_+$ variation in the vicinity of $r_{c2}$. The curves are drawn with $d=10$, $k=2$ (solid) and $d=10$, $k=3$ (dashed). In both cases, $Q=0.235$}
\end{center}
\end{figure}

In all dimensions $d$ and for all orders $k$, we find that there exists a range of values for the parameter $Q$, for which the function $T(r_+)$ has a couple of turning points. Since $C_p(r_+)=\dfrac{\partial_{r_+} M}{\partial_{r_+} T}$, we expect to find two points of divergence when we plot $C_p$ against $r_+$, indicating points at which $C_p$ changes sign discontinuously, signaling transitions between stable (+ve value for $C_p$) and unstable (-ve values for $C_p$) phases. This indeed turns out to be the case. We name the two turning points of $T(r_+)$ as $r_{c1}$ (the maximum) and $r_{c2}$ (the minimum). Samples of the typical variation of $C_p$ with $r_+$ in the vicinity of $r_{c1}$ and $r_{c2}$, for one particular set of values for the parameters $d$ and $Q$ with different values for $k$, are plotted in Figs. \ref{fig:C_vs_rplus_rc1_Q_0p235_d10_k_2_solid_k_3_dashed} and \ref{fig:C_vs_rplus_rc2_Q_0p235_d10_k_2_solid_k_3_dashed} respectively. From the analysis of the plots of $C_p$ against $r_+$ for various combinations of $d$, $k$ and $Q$, we observe that, when $r_+$ decreases, the transition at $r_{c2}$ is always from a stable phase to an unstable phase, whereas the nature of the transition at $r_{c1}$ changes from case to case, depending on the values of $d$, $k$ and $Q$. For example, it is clear from Fig. \ref{fig:C_vs_rplus_rc1_Q_0p235_d10_k_2_solid_k_3_dashed} that the second order theory (solid curve) in ten dimensions predicts an stable-to-unstable transition, while the third order theory (dashed curve) predicts a unstable-to-stable transition at $r_{c1}$ when $r_+$ decreases. In those cases where the transition at $r_{c1}$ is from a stable phase to an unstable phase, such as the one depicted by the solid curve in Fig. \ref{fig:C_vs_rplus_rc1_Q_0p235_d10_k_2_solid_k_3_dashed}, there obviously occurs a continuous sign-change in $C_p(r_+)$, as clearly seen in the figure.

\paragraph{}

\begin{figure}
\noindent\begin{minipage}{\textwidth}
\begin{minipage}{.4\textwidth}
 \includegraphics[scale=0.90]{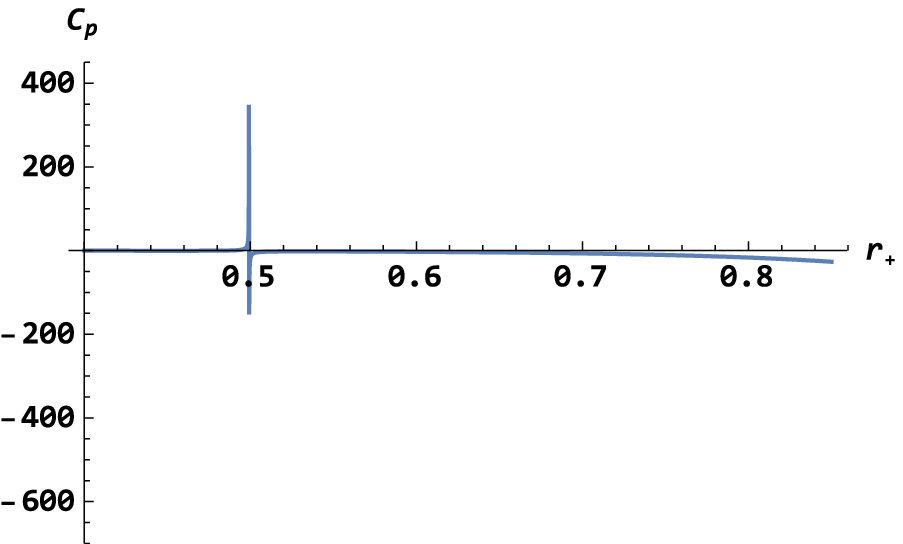}
\caption{\label{fig:C_vs_rplus_rc1_Q_0p235_d7_k_2}Divergence of $C_p(r_+)$ at $r_{c1}$, with $d=7$, $k=2$ and $Q=0.235$}
\end{minipage}%
\hfill
\begin{minipage}{.4\textwidth}
\includegraphics[scale=0.45]{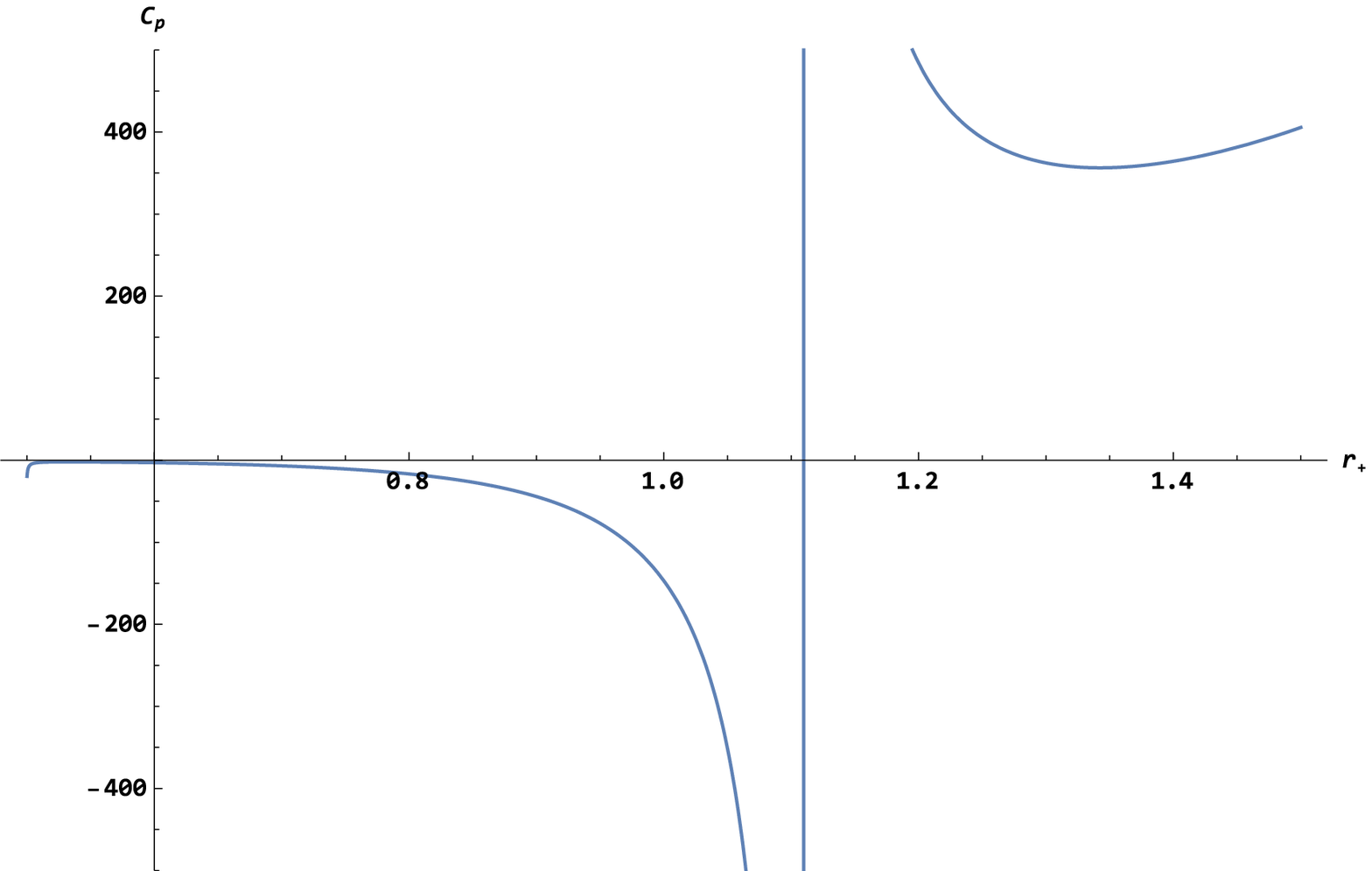}
\caption{\label{fig:C_vs_rplus_rc2_Q_0p235_d7_k_2}Divergence of $C_p(r_+)$ at $r_{c2}$, with $d=7$, $k=2$ and $Q=0.235$}
\end{minipage}
\end{minipage}\\~\\~\\
\begin{minipage}{\textwidth}
\begin{center}
\includegraphics[scale=1.0]{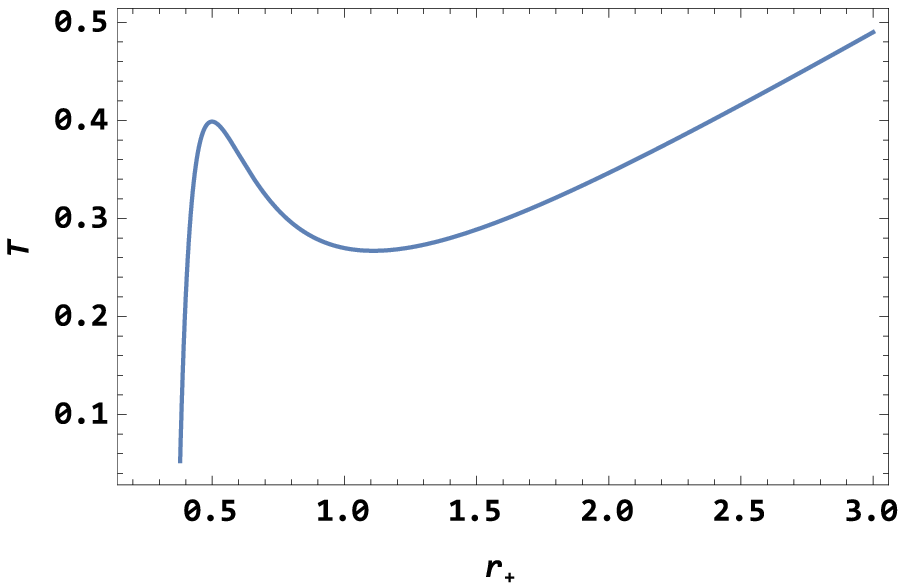}
\caption{\label{fig:T_vs_r_plus_d7k2Q_0p235}Variation of $T$ with $r_+$ for $d=7$, $k=2$ and $Q=0.235$}
\end{center}
\end{minipage}
\end{figure}

We now take another case and analyze the thermodynamic behaviour in some more detail. We select a black hole spacetime with $d=7,~k=2$ and $Q=0.235$. Plots of $C_p$ against $r_+$ in the neighborhood of $r_{c1}$ and $r_{c2}$ are given in Fig. \ref{fig:C_vs_rplus_rc1_Q_0p235_d7_k_2} and  \ref{fig:C_vs_rplus_rc2_Q_0p235_d7_k_2}. From the figures, it is obvious that, for this particular case, as $r_+$ decreases, the transition at $r_{c2}$ is from a stable phase to an unstable phase, whereas the transition at $r_{c1}$ is from an unstable phase to a stable one.

\paragraph{}

Let us try to analyze the thermodynamic stability of the spacetime using  Figs. \ref{fig:C_vs_rplus_rc1_Q_0p235_d7_k_2}-\ref{fig:T_vs_r_plus_d7k2Q_0p235}. We will see that the thermodynamic stability depends on both the size of the black hole and the temperature of the background AdS spacetime. The thermodynamic behaviour of the black holes depends crucially on whether the temperature ($T_0$) of the background spacetime (also called thermal bath) is \textbf{(i)} larger than the local maximum value ($T_{max}$) of $T(r_+)$, \textbf{(ii)} between $T_{max}$ and the local minimum value ($T_{min}$) of $T(r_+)$, or \textbf{(iii)} lower than $T_{min}$. Another important factor that determines the thermodynamic behaviour in all these three cases is the size $r_+$ of the black hole itself - whether it is greater than $r_{c2}$, in between $r_{c2}$ and $r_{c1}$ or less than $r_{c1}$. We analyze some of the possible scenarios here:

\paragraph{Case (i) - $T_0>T_{max}$:\\~\\}

In this case, we see that, similar to the case of Schwarzschild-AdS black holes, very large black holes in the model can always attain equilibrium with an external thermal bath at a finite temperature, since the specific heat is positive in this region. There is one difference though - when $r_+$ is large and the temperature $T_0$ of the bath is higher than the maximum value of $T(r_+)$, a straight line parallel to the horizontal axis meets the $T(r_+)~-~r_+$ curve at only one point, which means that there exists only one final, stable, equilibrium configuration for such a black hole at that background temperature, with a horizon radius which is larger than $r_{c2}$. This is not the case for uncharged black holes - the uncharged spacetime always has two equilibrium configurations - one being unstable and the other being stable - as long as the temperature of the bath is higher than the minimum of $T(r_+)$\cite{bhscan}. Thus, for the charged case, a black hole with $r_+>r_{c2}$ will get drawn towards the equilibrium state at temperature $T_0$, since the specific heat is positive for $r_+>r_{c2}$.

\paragraph{Case (ii) - $T_0<T_{min}$:\\~\\}

Charged AdS Lovelock black holes can attain equilibrium with a thermal bath of \emph{any} positive temperature, whereas the uncharged ones are known\cite{bhscan} to be unable to attain equilibrium with a bath of temperature lower than the minimum of $T(r_+)$. Case \textbf{(ii)} is an example of such a scenario. Here again, there exists only one thermodynamically stable equilibrium configuration that the small black holes can get drawn towards, since the straight line parallel to the horizontal axis still cuts the $T(r_+)~-~r_+$ curve at only one point. In this case, the horizon radius for the equilibrium state will be smaller than $r_{c1}$. The specific heat is positive in the region $r_+<r_{c1}$, so that a black hole with $r_+<r_{c1}$ tends to make a transition towards this equilibrium state rather than away from it. 

\paragraph{Case (iii) - $T_{min}<T_0<T_{max}$:\\~\\}

It is clear from Fig. \ref{fig:T_vs_r_plus_d7k2Q_0p235} that, in this range, each value of $T_0$ corresponds to three equilibrium states of different radii - say, $r_{s1}~(\textmd{ less than }r_{c1}$, corresponding to a locally stable state), $r_u(\textmd{ in between }r_{c1} \textmd{ and }r_{c2}$, corresponding to an unstable state) and $r_{s2}~(\textmd{ larger than }r_{c2}$, corresponding to a locally stable state). The points $r_{s1}$ and $r_{s2}$ exist in regions with positive specific heat, so that initial black hole states with $r_+<r_{c1}$ and $r_+>r_{c2}$ are drawn towards these equilibrium points respectively. On the other hand, the point $r_u$ exists in a region with negative specific heat, so that initial black hole states with $r_{c1}<r_+<r_{c2}$ are drawn away from this equilibrium point. Thus, if the temperature of the thermal bath falls in the range $T_{min}<T_0<T_{max}$, the resultant thermodynamic behaviour of the black hole will depend on its initial size. Essentially, initial black hole states with $r_+>r_u$ will evolve towards an equilibrium configuration with $r_+=r_{s2}$ and those with $r_+<r_u$ will tend to evolve towards a configuration with $r_+=r_{s1}$. This is in contrast with the uncharged case, where a black hole with an initial size $r_+<r_u$ can never reach equilibrium\cite{bhscan}.

\paragraph{}

\begin{figure}[H]
\begin{center}
\includegraphics[scale=1.0]{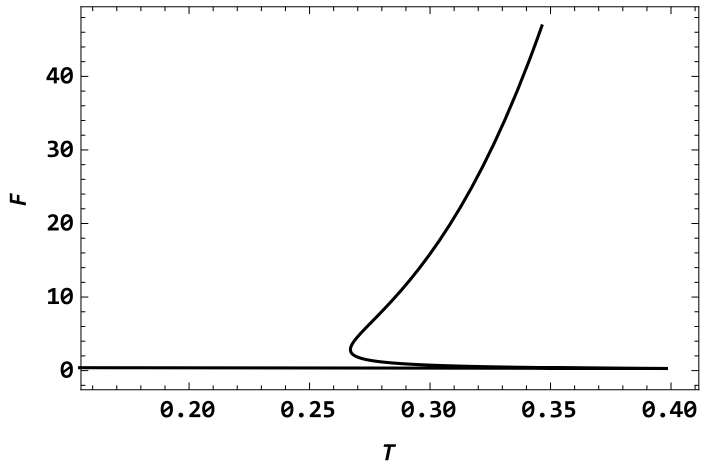}
\caption{\label{fig:F_vs_rplus_Q_0p235_d7_k_2}Free energy of the charged black hole as a function of $r_+$ for $d=7$, $k=2$ and $Q=0.235$}
\end{center}
\end{figure}

The occurrence of a phase transition in this sample case is also indicated by a plot between the Gibbs free energy $F(r_+)=M(r_+)-T(r_+)S(r_+)$ and the horizon temperature $T$, given in Fig. \ref{fig:F_vs_rplus_Q_0p235_d7_k_2}. The presence of a cusp in the plot indicates that there occurs a second order phase transition in the black hole spacetime.

\section{Geometrothermodynamic (GTD) Analysis}\label{sect:gtd}

We employ the Legendre invariant method of Quevedo\cite{quevedo1,quevedo2,quevedo3,quevedo4,bravetti1,bravetti2} in order to study the phase transition in the geometric formalism. We choose the entropy representation, in which the Ricci scalar $R_R$ that represents the thermodynamic interaction of the system is derived from a thermodynamic metric which is defined in terms of the second derivatives of the entropy $S$ of the system, considered as a function of the relevant extensive parameters. For our spacetime representing the charged, AdS Lovelock black holes, we take $M$ and $Q$ as the extensive parameters. Since $M$ can conveniently be expressed as a function of the horizon radius $r_+$, we take $M=M(r_+)$, $Q=Q(r_+)$, $S=S(r_+)$ and compute the Ricci scalar $R_R$ as a function of $r_+$. Identifying ${M,Q}$ as the set of extensive thermodynamic variables and $S(M,Q)$ as the thermodynamic potential $\Phi$ of the system, the Legendre invariant thermodynamic metric $g^Q$ is computed using the relation,

\begin{equation}\label{quevedo_metric}
g^Q=\left(
\begin{array}{cc}
g_{11}&0\\0&g_{22}
\end{array}
\right)~,
\end{equation} 

where $g_{11}$ and $g_{22}$ are given by,

\begin{eqnarray}
g_{11}&=&-\left(M\dfrac{\partial_r S}{\partial_r M}+Q\dfrac{\partial_r S}{\partial_r Q}\right)\left(\dfrac{\partial_r M ~ \partial_{rr}{S}-\partial_r S ~ \partial_{rr}{M}}{(\partial_r M)^3}\right) \textmd{, and,}\label{g11}\\
g_{22}&=&\left(M\dfrac{\partial_r S}{\partial_r M}+Q\dfrac{\partial_r S}{\partial_r Q}\right)\left(\dfrac{\partial_r Q ~ \partial_{rr}{S}-\partial_r S ~ \partial_{rr}{Q}}{(\partial_r Q)^3}\right).\label{g22}
\end{eqnarray} 

\begin{figure}[H]
\noindent\begin{minipage}{\textwidth}
\begin{minipage}{.4\textwidth}
 \includegraphics[scale=0.35]{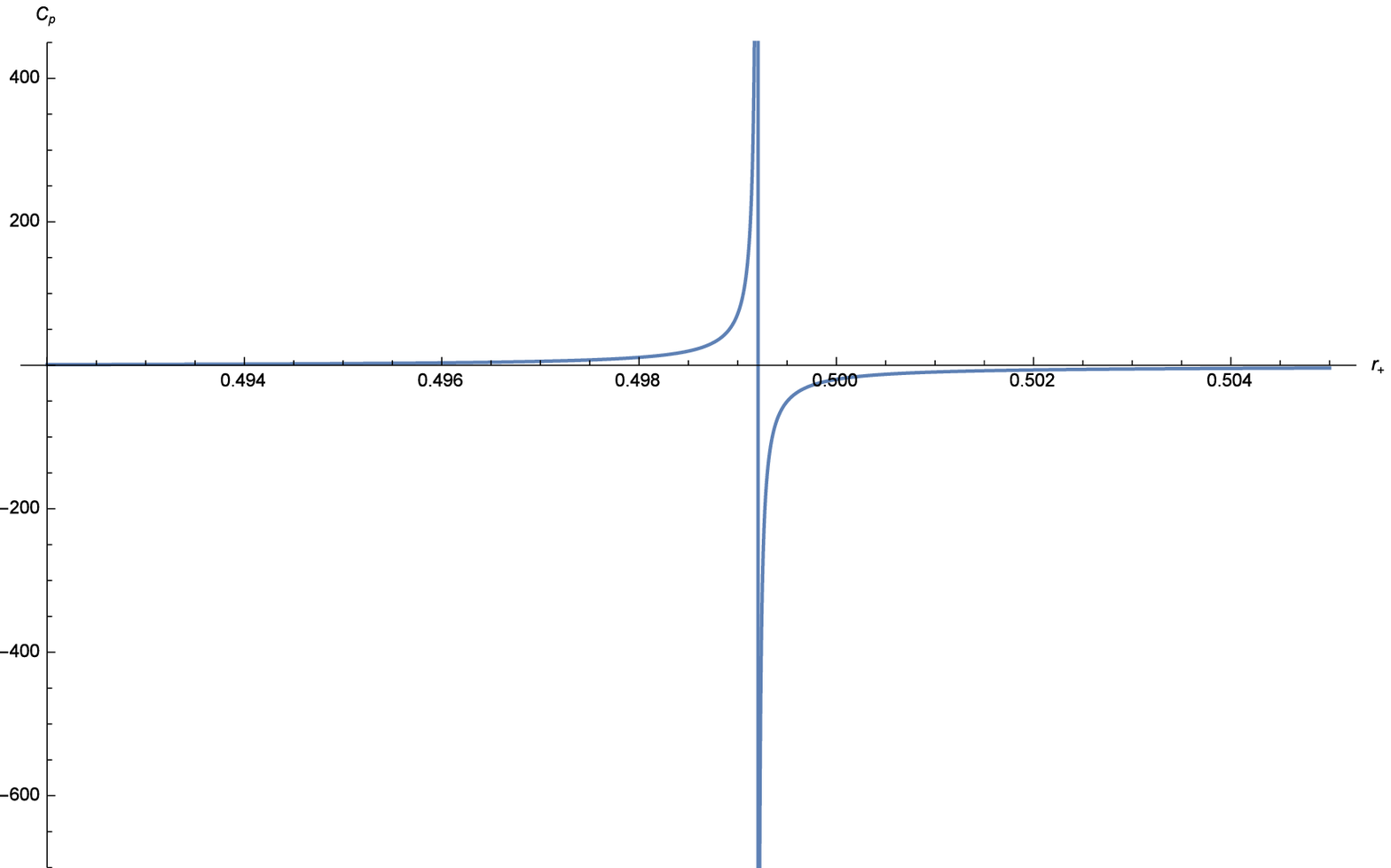}
\caption{\label{fig:C_p_vs_r_plus_d7k2Q_0p235_first_point_of_divergence}Divergence of $C_p(r_+)$ at $r_{c1}$. Here, $d=7$, $k=2$ and $Q=0.235$}
\end{minipage}
\hfill
\begin{minipage}{.4\textwidth}
\includegraphics[scale=0.35]{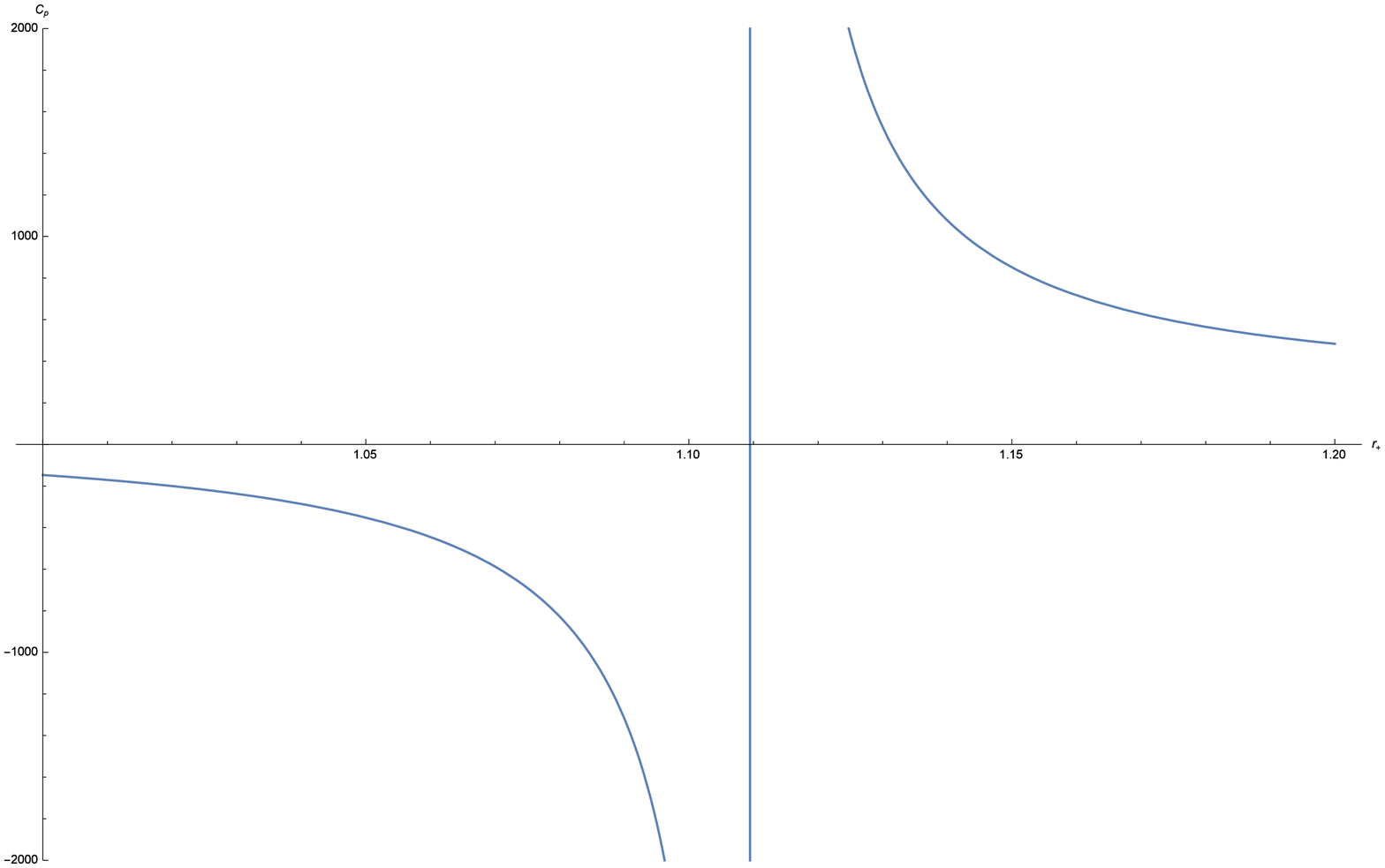}
\caption{\label{fig:C_p_vs_r_plus_d7k2Q_0p235_second_point_of_divergence}Divergence of $C_p(r_+)$ at $r_{c2}$. Here, $d=7$, $k=2$ and $Q=0.235$}
\end{minipage}
\end{minipage}
\end{figure}

\begin{figure}[H]
\noindent\begin{minipage}{\textwidth}
\begin{minipage}{.4\textwidth}
\includegraphics[scale=0.35]{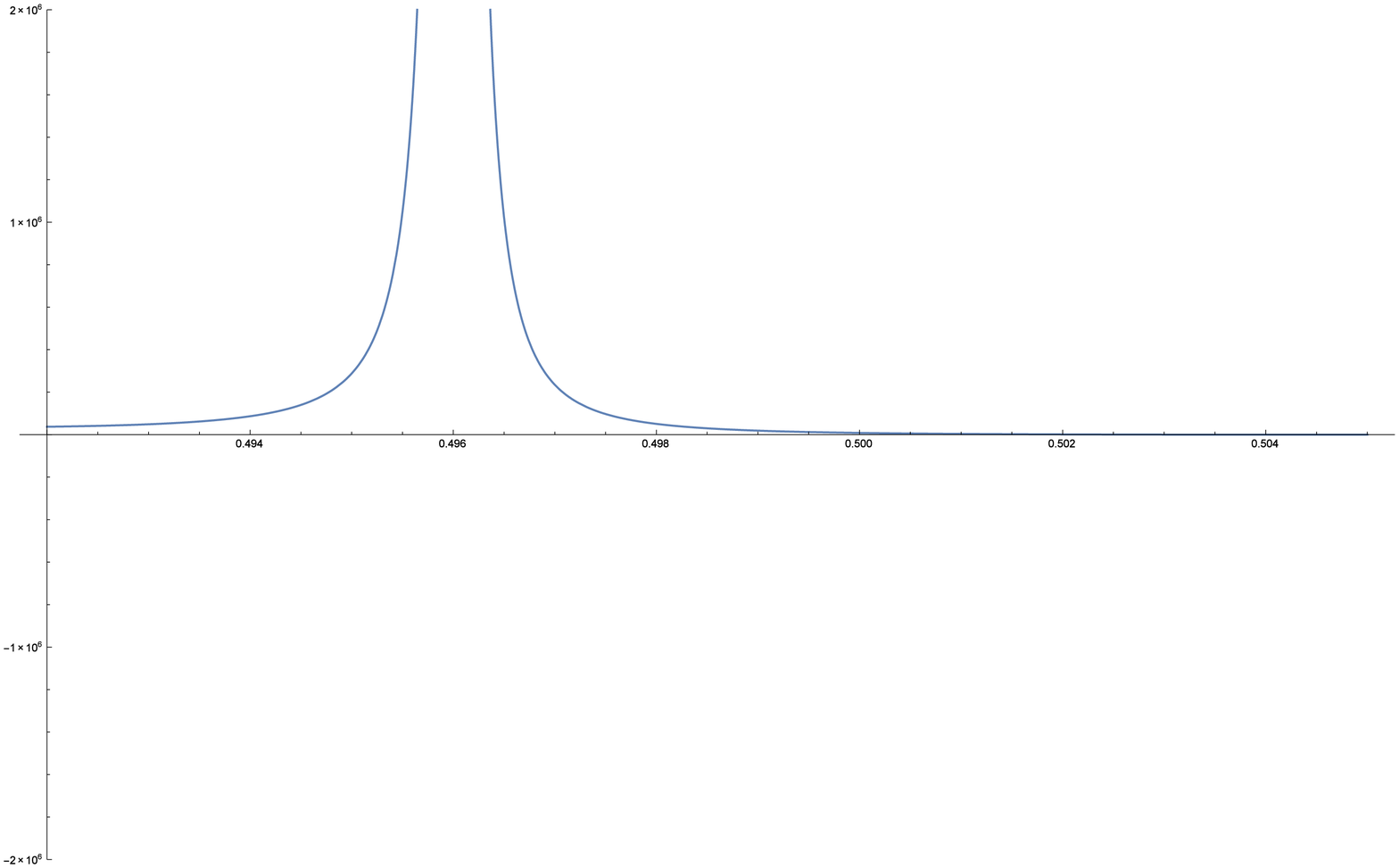}
\caption{\label{fig:gtd_curvature_plot_entropy_rep_d7k2Q_0p235_first_divergence}Divergence of $R_R$ near $r_{c1}$. Here, $d=7$, $k=2$ and $Q=0.235$}
\end{minipage}
\hfill
\begin{minipage}{.4\textwidth}
\includegraphics[scale=0.35]{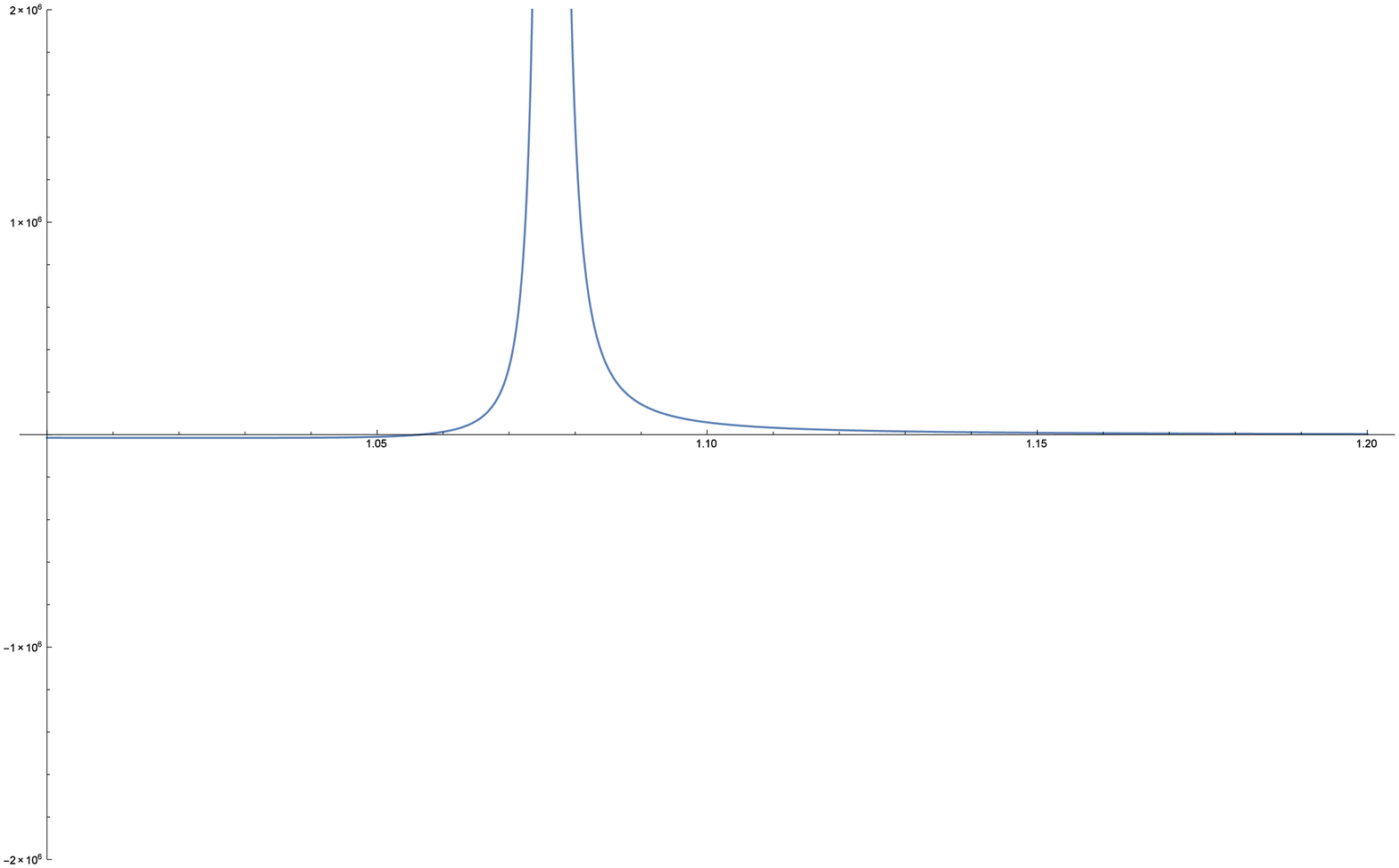}
\caption{\label{fig:gtd_curvature_plot_entropy_rep_d7k2Q_0p235_second_divergence}Divergence of $R_R$ near $r_{c2}$. Here, $d=7$, $k=2$ and $Q=0.235$}
\end{minipage}
\end{minipage}
\end{figure}

Note that the symbol $r$ replaces $r_+$ in (\ref{g11}) and (\ref{g22}). Although the analytic calculation is straightforward, the resultant expressions for the metric-components and that of the Ricci scalar $R_R$ are too long to be explicitly included here. Therefore, we resort to numerical analysis and study the behaviour of $R_R$ graphically as a function of $r_+$. In Figs. \ref{fig:C_p_vs_r_plus_d7k2Q_0p235_first_point_of_divergence}-\ref{fig:gtd_curvature_plot_entropy_rep_d7k2Q_0p235_second_divergence}, we plot $R_R(M,Q)$ against $r_+$ for specific values of the black hole parameters and compare it with the corresponding plots of the heat capacity $C_p$, also plotted against $r_+$ in exactly the same range. From the plots, it is clear that the divergences in $R_R$ occur at points which are very near to those at which the heat capacity diverges. Hence, we conclude that the usual thermodynamic approach and the GTD method are in agreement in predicting the thermodynamic behaviour of the black hole spacetime.

\section{Area Spectrum of large charged AdS Black holes in Lovelock Model}\label{sect:area}

In this section, we compute the horizon area spectrum of large ($r_+ \gg R$) charged, AdS black holes in Lovelock model. The fact that horizon area of black holes is quantized was proposed for the first time by Bekenstein\cite{bekenstein1,bekenstein2,bekenstein3}. He found that the horizon area of a non-extremal black hole is a classical adiabatic invariant. It is known from field theory (Ehrenfest Principle) that the presence of a periodicity in the classical theory of a system points to the existence of an adiabatic invariant with a discrete spectrum in the corresponding quantum theory. We follow the recent proposal by Majhi and Vagenas\cite{majhi} that the quantity $I=\displaystyle \sum \int p_i dq_i$ can be taken as the classical adiabatic quantity in the case of black hole spacetimes, where $q_i$ and $p_i$ are conjugate variables describing the dynamics of the system.

\paragraph{}

For a spacetime whose metric is given by 

\begin{equation}\label{adiabatic_metric}
ds^2=f(r)d \tau^2+\dfrac{1}{f(r)}dr^2+r^2d\Omega_{d-2}^2\ ,
\end{equation} 

where $\tau=i t$ is the Euclidean time coordinate, we take $q_0=\tau$ and $q_1=r_h$ as the dynamical variables of the system and consider the Hamiltonian $H$ as a function of $q_i$ and $p_i$. Then, taking into account one of the Hamiltonian equations of motion of the system, namely, $p_i=\dfrac{\partial H}{\partial \dot{q_i}}$, it is possible to show that the adiabatic invariant $I$ for the spacetime takes the form,

\begin{eqnarray}\label{adia_1}
I&=&-2 i \int \int_0^H \dfrac{dH'}{f(r)}dr.
\end{eqnarray} 

Near the horizon $r_h$, we can approximate $f(r)\approx \kappa(r-r_h)$, where $\kappa=\left.\dfrac{df}{dr}\right|_{r=r_h}$. Also, the temperature $T$ of the horizon is given by $T=\dfrac{1}{4 \pi}\kappa$. Substituting all these into (\ref{adia_1}), we get,

\begin{equation}\label{adiabatic_hamil}
I=\dfrac{1}{2}\int_0^H\dfrac{dH'}{T}.
\end{equation} 

Considering the black hole spacetime as a thermodynamic system with extensive variables $S$ and $Q$, we equate the Hamiltonian $H'$ to the mass $M$ of the black hole, so that the first law of thermodynamics reads

\begin{equation}\label{first_law}
dH'=dM=T dS+\Phi dQ,
\end{equation} 

$\Phi=\dfrac{\partial M}{\partial Q}$ being the electric potential. Thus, (\ref{adiabatic_hamil}) gives

\begin{equation}\label{adiabatic_final}
I=\dfrac{1}{2}(S+\int_0^Q \dfrac{\Phi}{T}\ dQ').
\end{equation} 

The temperature $T$ of the horizon is given by $T=\dfrac{1}{4\pi}\left.\dfrac{df}{dr}\right|_{r=r_h}$, and $\Phi=\dfrac{\partial M}{\partial Q'}$, where $M(Q')$ by replacing $Q$ with $Q'$ in (\ref{mass}). We get

\begin{equation}\label{electro_pot}
\Phi=\dfrac{\epsilon r_h^{3-d}}{d-3}\ Q'.
\end{equation} 


We compute the second term on the RHS of (\ref{adiabatic_final}) after substituting the values of $\Phi$ and $T$. Its value turns out to be, 

\begin{equation}\label{second_term}
\int_0^Q \dfrac{\Phi}{T}\ dQ'=\dfrac{a}{2c}\ \ln (b-c\ Q^2) \textmd{, where}
\end{equation} 

\begin{equation*}
a=\dfrac{\epsilon r_h^{3-d}}{d-3},
\end{equation*}

\begin{equation*}
b=\dfrac{1}{4\pi}\Biggl[\dfrac{2r_h}{R^2}+\dfrac{2G_kM(d-2k-1)}{kr_h^{d-2k}}\left(1+\dfrac{r_h^2}{R^2}\right)^{1-k}\Biggr] \textmd{, and}
\end{equation*} 

\begin{equation*}
c=\dfrac{1}{4\pi}\Biggl[\dfrac{2\epsilon G_k(d-k-2)}{k(d-3)r_h^{2(d-k)-3}}\left(1+\dfrac{r_h^2}{R^2}\right)^{1-k}\Biggr].
\end{equation*}

Thus, (\ref{adiabatic_final}) gives,

\begin{equation*}
S=2I-\dfrac{a}{2c}\ \ln (b-c\ Q^2)
\end{equation*} 

According to Bohr-Sommerfeld quantization condition, $I=n \hbar$, so that we can write

\begin{equation}\label{quantized_entropy_expression}
S=2n \hbar-\dfrac{a}{2c}\ \ln (b-c\ Q^2)
\end{equation}

Now, it is to be noted that the usual area law, namely $S \propto A$, is unique to first order theories of gravity like the General Theory of Relativity, for which the order parameter $k=1$. In general, the area law is not followed by black hole spacetimes in theories where $k \neq 1$. However, the area law can approximately be recovered in these theories if we restrict our attention to very large black holes, i.e. those with $r_+ \gg R$. For such black holes, the general expression for the entropy for $S(r_+)$, given by (\ref{entropy}), reduces to

\begin{equation}\label{entropy_approx_1}
S(r_+) \approx k\ \left(\dfrac{2 \pi \kappa_B}{(d-2)G_kR^{2(k-1)}}\right)\ r_+^{d-2},
\end{equation} 

which can be written in terms of the area $A$ of the event horizon as

\begin{equation}\label{entropy_approx_2}
S(r_+) \approx k\ \left(\dfrac{2 \pi \kappa_B}{(d-2)G_kR^{2(k-1)}}\right)\ \left(\dfrac{A}{\Omega_{d-2}}\right).
\end{equation}

Using (\ref{entropy_approx_2}) in (\ref{quantized_entropy_expression}), we can write an expression for the quantized area of large, charged black holes in the asymptotically AdS Lovelock model as,

\begin{equation}\label{quantized_area}
A=\gamma \left( n\hbar - \dfrac{a}{4c} \ln\ (b-c\ Q^2)\right),
\end{equation} 

where the constant $\gamma$ is given by,

\begin{equation*}
\gamma=\dfrac{(d-2) \Omega_{d-2}G_kR^{2(k-1)}}{\pi \kappa_B}.
\end{equation*}

Thus the horizon-area of large, charged, black holes in the model turns out to be quantized, with a logarithmic correction term added to it. The dependence of the area-quantum on the order of the theory $k$ and the AdS radius $R$ is evident from the expression for $\gamma$. It is to be noted that the dependence on $R$ becomes evident only when one considers theories with $k \neq 1$. It is interesting to note that the logarithmic correction term itself does not turn out to be quantized.

\section{Conclusions}\label{sect:conclusion}

In this paper, we considered the thermodynamic behaviour of charged, asymptotically AdS and spherically symmetric black hole solutions of the Lovelock model of gravity, where the higher-order coupling constants are chosen so as to make the AdS radius $R$ equal for all orders. The main objective has been to investigate the thermodynamic stability of such black holes and to look for possible phase transitions between various black hole states. Two approaches were adopted toward that end - (1): the usual thermodynamic approach in which one computes the specific heat of the spacetime and looks for divergences which signal the occurrence of second order phase transitions between various states, and (2): the method of geometrothermodynamics, in which one studies the thermodynamic interaction of the black hole by applying methods of differential geometry to the thermodynamic phase space of the system. 

\paragraph{}

Using the usual methods of black hole thermodynamics, we calculated different thermodynamic parameters of the system such as the horizon temperature, entropy and the specific heat. We found that the horizon temperature $T$, when written as a function of the horizon radius $r_+$, has a couple of turning points, compared to just one in the uncharged case. Entropy $S$ happens to be a monotonic function of $r_+$, while the specific heat $C_p$ exhibits divergence at two points corresponding to the turning points of $T(r_+)$. From the plots of $T(r_+)$ and $C_p(r_+)$ against $r_+$, we were able to deduce the thermodynamic behaviour of the black holes. 

\paragraph{}

We found that large black holes are always able to attain thermodynamic equilibrium with the background AdS spacetime (the thermal bath) as long as the bath has a temperature greater than the local maximum value of $T(r_+)$. In this case, there exists only one stable equilibrium configuration for a black hole at any bath temperature $T_0$. A similar conclusion can be arrived at in the case of small black holes placed inside a bath at a temperature $T_0$ that is less than the local minimum value of $T(r_+)$. There exists a stable equilibrium configuration in this case as well, in contrast with the case of uncharged black holes. In this case, as in the previous case, initial black hole states get drawn towards the respective equilibrium configurations, since the specific heat is positive in both cases. When the temperature of the bath is in between the local maximum and local minimum values of $T(r_+)$, each bath temperature $T_0$ corresponds to three equilibrium black hole configurations - two stable states and one unstable state. In such a case, we concluded that initial black hole configurations would be drawn towards one of the stable points, depending on their initial size.

\paragraph{}

In order to perform the geometric analysis of the thermodynamic evolution, we followed the method of geometrothermodynamics of Quevedo\cite{quevedo1,quevedo2}. We chose what is known as the entropy representation of the thermodynamic phase space and computed the scalar curvature $R_R$ derived from a Legendre invariant thermodynamic metric, the components of which are calculated using the second derivatives of the entropy of the system. We chose the mass and the charge of the black hole as the extensive parameters. Since the expressions for the metric components and the scalar curvature were too long to treat analytically, we resorted to the graphical method, plotting the scalar curvature as a function of the horizon radius $r_+$. From the plot, we found that the scalar curvature diverges at points that are very close to the points of divergence of the specific heat of the black hole, indicating that the thermodynamic phase transitions of the black hole correspond to the singularities in the corresponding thermodynamic phase space. Thus, the results of geometrothermodynamics were found to be in agreement with those of ordinary black hole thermodynamics.

\paragraph{}

Next, we computed the horizon area spectrum of large, charged AdS black holes in the model, motivated mainly by the AdS/CFT correspondence. We computed the adiabatic invariant $\displaystyle \sum \int p_i dq_i $ for the black hole spacetime taking the Euclidean time $\tau=it$ and the horizon radius $r_h$ as the dynamical variables. The first law of thermodynamics was made use of during the computation. We applied the Bohr-Sommerfeld quantization rule to the invariant and found that the entropy is a quantized entity with a logarithmic correction term added to it. In the limit of large black holes, the entropy becomes proportional to the horizon area for higher order Lovelock theories and we were able to show that the area of such black holes also can be written as a quantized number with a logarithmic correction term added to it. The spacing $\gamma$ between the various quanta was found to be dependent on the order $k$ of the theory (and by extension the dimension $d$ of the spacetime) and the value of the AdS radius $R$, although this dependence become evident only in higher dimensions and higher order theories.

\section{Acknowledgments}

CBP would like to acknowledge financial assistance from UGC, India through the UGC-RFSMS Scheme. VCK and JS would like to acknowledge financial assistance from UGC through a major research project. VCK  would like to acknowledge Associateship of IUCAA, Pune, India.

\end{document}